\documentclass[aip,apl,reprint]{revtex4-1}
\usepackage{graphicx}
\usepackage{graphics}

\begin{document}

\title{Local injection of pure spin current generates electric current
vortices}

\author{Ya.\ B. Bazaliy}
  \affiliation{University of South Carolina, Columbia SC 29208, USA}
  \email{yar@physics.sc.edu}
\author{R. R. Ramazashvili}
 \affiliation{Laboratoire de Physique Th\'eorique, IRSAMC,
 Universit\'e de Toulouse, CNRS, 31062 Toulouse, France}
 \email{revaz@irsamc.ups-tlse.fr}
\date{\today}

\begin{abstract}
We show that local injection of pure spin current into an electrically disconnected
ferromagnetic--normal-metal sandwich induces electric currents, that run along
closed loops inside the device, and are powered by the source of the spin injection.
Such electric currents may significantly modify voltage distribution in spin-injection
devices and induce long-range tails of spin accumulation.
\end{abstract}

\maketitle

\begin{figure}[b]
\center
\includegraphics[width = 0.45\textwidth]{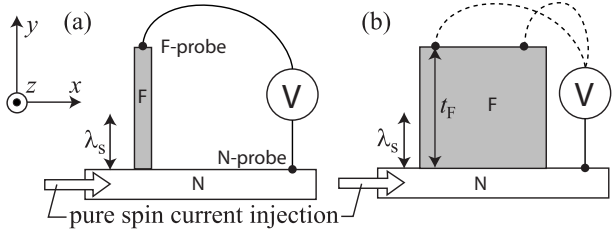}
\caption{Electrically disconnected device with pure spin current injected
into the N-layer from the side; (a) narrow F-electrode (b) wide F-electrode
with two different F-probe positions leading to different results for the $V$.}
 \label{fig:device}
\end{figure}

Injection of pure spin current and its subsequent manipulation in spintronic devices
\cite{maekawa_book2012} has been viewed as a milestone in
realization of ``spin electronics'', where electron spin would be
carrying signal on a par with the charge.
In classic experiments of Johnson and Silsbee \cite{johnson-silsbee_prl1985},
pure spin current was injected into an electrically disconnected device.
Since there was no electric current ${\bf j}$ entering or leaving the device,
it was tacitly assumed that ${\bf j}$ should also be zero everywhere inside it.
Johnson and Silsbee found that spin current injection nevertheless generates
a voltage $V$ between the ferromagnetic (F) and normal (N)
elements (Fig.~\ref{fig:device}). In a diffusive transport regime,
where electron momentum relaxes much faster than its spin, such a voltage can
be described in terms of the ``Valet-Fert model'',\cite{vanson_prl1987,
valet-fert_prb1993, takahashi_prb2003, rashba_prbrc2000, rashba_epjb2002} outlined below. Johnson and Silsbee
\cite{johnson-silsbee_prl1985} predicted the $V$ to be proportional to the
spin accumulation at the F/N boundary, and independent of the measuring probe
position  --- as long as the electric current was absent, and the F-probe was
placed at a point where spin accumulation has relaxed to zero (i.e., further
than several spin relaxation lengths $\lambda_s$
away from the F/N boundary).

This statement is true and transparent in the case
of a narrow F/N contact (Fig.~\ref{fig:device}a).
However, if the contact is wide enough for the spin accumulation
to vary substantially along the F/N interface Fig.~\ref{fig:device}b), then it is not
clear which accumulation value should be used in the Johnson-Silsbee formula.
This practical issue was investigated, e.g., in the the Ref.~\onlinecite{hamrle_prb2005},
where it was found that $V$ does depend on the probe position, even if the
thickness $t_F$ of the F layer exceeds $\lambda_s$ (Fig.~\ref{fig:device}).

On the one hand, emergence of a non-uniform vol\-ta\-ge in a system with
non-uniform spin accumulation appears to be natural. On the other hand, a
potential gradient in the F region with vanishing non-equilibrium spin
accumulation can only mean the presence of electric current. How does this
correspond to the absence of ${\bf j}$ in the Johnson-Silsbee picture?
Here we show that even if electric currents do not enter the
device,\cite{note} they are still induced inside it. These
internal currents circulate along closed loops that cross
the F/N interface, and are maintained by the external source that produces
the pure-spin injection. We demonstrate the existence of electric current loops
and study their influence on the voltage and spin accumulation distributions.
Current loops are akin to Eddy currents generated by oscillating magnetic fields,
except that the present phenomenon occurs in a non-equilibrium steady state.
We show that such electric vortices are not limited to spin transport,
and shall be expected whenever electric current is coupled to
another diffusive current by linear relationships with Onsager cross-coefficients.

We will consider setups with collinear magnetization. As detailed in
Ref.~\onlinecite{rashba_epjb2002}, in the Valet-Fert model carrier
distributions for spin $\alpha = \, \uparrow,\downarrow$ are characterized by
different electrochemical potentials $\mu_{\alpha}$. With two
conductivities $\sigma_{\uparrow, \downarrow}$ being different in a
ferromagnet, the currents\cite{footnote_number_currents} %${\bf j}_{\alpha}$
carried by the two spin populations are given by ${\bf j}_{\alpha} = - (\sigma_{\alpha}/e^2) \nabla \mu_{\alpha}$.
Conservation of electric current ($\partial_t n + {\rm div} {\bf j} = 0$)
and spontaneous relaxation of spin ($\partial_t n^s + {\rm div} {\bf j}^s = - n^s / \tau_s$)
yield steady-state equations
\begin{equation}\label{eq:divj_divjs}
{\rm div}{\bf j} = 0, \quad
    {\rm div}{\bf j}^s = - n_s/\tau_s \ ,
\end{equation}
where ${\bf j} = {\bf j}_{\uparrow} + {\bf j}_{\downarrow}$  and ${\bf j}^s =
{\bf j}_{\uparrow} - {\bf j}_{\downarrow}$ are electric and spin currents,
the $n$ and $n_s$ are the non-equilibrium charge density
and spin accumulation, and $\tau_s$ is the spin relaxation time. The average
potential $\mu = (\mu_{\uparrow} + \mu_{\downarrow})/2$ is the quantity
measured by an ideal voltmeter, while the spin potential
$\mu^s~=~\mu_{\uparrow}~-~\mu_{\downarrow}$ characterizes the non-equilibrium
spin accumulation. The currents ${\bf j}$ and ${\bf j}^s$ can be written as
\begin{eqnarray}
\label{eq:j}
{\bf j} &=& - \frac{\sigma}{e^2} \, (\nabla\mu + \frac{p}{2}\nabla\mu^s)
\\
\label{eq:js}
{\bf j}^s &=& -\frac{\sigma}{2 e^2} \, (\nabla\mu^s + 2 p \nabla\mu)
\end{eqnarray}
with $\sigma = \sigma_{\uparrow} + \sigma_{\downarrow}$, and the polarization
$p~=~(\sigma_{\uparrow}~-~\sigma_{\downarrow})/\sigma$.
Note that in the Eqs. (\ref{eq:j}-\ref{eq:js}) spin and charge are coupled by $p \neq 0$.
We will assume $\sigma$, $p$ and $\tau_s$ to be piecewise
constant, undergoing jumps at interfaces between different materials.
Within each uniform region, Eqs.~(\ref{eq:divj_divjs}-\ref{eq:js}) yield
\begin{equation}
\label{eq:bulk_barmu_mus}
\Delta\mu = - \frac{p}{2} \Delta\mu^s \ , \quad
\lambda_s^2 \Delta\mu^s = \mu^s \ ,
\end{equation}
with $\lambda_s$ being the spin relaxation length.\cite{rashba_epjb2002}
The interfaces will be assumed transparent (continuity of $\mu$ and $\mu^s$)
and spin-inactive (continuity of the $j^s_{\perp}$ component, normal to the
boundary).

\begin{figure}[t]
\center
\includegraphics[width = 0.5\textwidth]{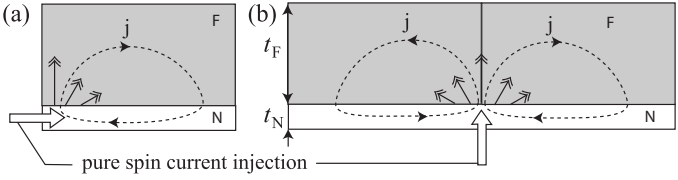}
    \caption{(a) Device with an extended F-layer. Pure spin current is injected
    into the N-layer from the side. Double arrows represent the effective EMF
    generated near the F/N boundary. Dashed line is the generated electric
    current loop; (b) ``Spin fountain'', i.e. symmetric extension of device (a),
    with pure spin current locally injected into the N-layer from below.
    For $t_N \ll \lambda_n$ the currents and potentials in (a) will be
    approaching those in the right half of (b).}
 \label{fig:model_geometry}
\end{figure}

First, we show that non-uniform spin accumulation near the F/N boundary
inevitably produces electric current, even if the latter is not injected from
outside. The Eq. (\ref{eq:j}) implies
\begin{equation}
\label{eq:curl_j}
{\rm curl} \left( \frac{\bf j}{\sigma/e^2} \right)
 = -\frac{1}{2} \nabla p \times \nabla\mu^s \ .
\end{equation}
Now, $p$ evolves from $p = 0$ in the normal metal to $p \neq 0$ in the
ferromagnet, thus $\nabla p \neq 0$. If $\nabla \mu^s$ has a component
perpendicular to $\nabla p$, i.e. if spin accumulation varies along the
interface between materials with different polarizations $p$, then ${\rm
curl} \left( e^2{\bf j}/{\sigma} \right) \neq 0$, and thus ${\bf j} \neq 0$.
Since current cannot cross an outer boundary of an electrically
disconnected device, it circulates inside, forming closed loops.
These loops cannot be confined to any of the uniform parts of the
device, and thus cross the boundaries between them. Indeed, the
presence of a current loop in a region of constant $\sigma$ and $p$
would mean ${\rm curl} \, {\bf j} \neq 0$, which is impossible due to Eq. (\ref{eq:j}):
in uniform regions, ${\bf j}$ is the gradient of a function. Thus, current
lines must form a vortex with the core somewhere at the F/N boundary.

The Eq.~(\ref{eq:j}) can also be interpreted as follows. Electric current is
driven by two forces: one is the conventional electrochemical potential
gradient, the other is an effective electromotive force (EMF)
${\cal E} = - p (\sigma/e^2) \nabla\mu^s / 2$ due to the non-equilibrium spin
accumulation $\mu^s$.\cite{rashba_epjb2002} Both the $\mu^s$ and its gradient
decay away from the spin current injection point, thus an EMF region appears
around it (Fig.~\ref{fig:model_geometry}(a)), producing the current loops.

Generation of electric current vortices is not limited to
spintronics. Consider coupled electric and heat transport
\begin{eqnarray}
\label{eq:current-heat}
{\bf j}_e &=& -\sigma \nabla\phi - S \sigma  \nabla T \ ,
 \\
\nonumber
{\bf q} &=& -\Pi \sigma \nabla\phi  - \kappa \nabla T \ ,
\end{eqnarray}
where ${\bf j}_e$ is the electric current, $\bf q$ is the heat flux, $\phi$ is the electric potential, $\kappa$ is the thermal conductivity, and $S$ and
$\Pi$ are Seebeck and Peltier coefficients. Si\-mi\-lar\-ly to how the Eq.
(\ref{eq:curl_j}) follows from the Eq. (\ref{eq:j}), the Eq.
(\ref{eq:current-heat}) implies that a temperature gradient satisfying
$\nabla S \times \nabla T \neq 0$ produces current loops at the interface
between materials with different Seebeck coefficients.

Now, we choose a symmetric device in Fig.~\ref{fig:model_geometry}(b) as a
simple setting to demonstrate the loop
current ge\-ne\-ration in a specific geometry. As the thickness $t_N$ of the
normal metal film decreases, we expect the spin accumulation to become ever
more uniform across the N-film. Then the solution for a realistic device with
pure spin current injected from the side as in
Fig.~\ref{fig:model_geometry}(a) will be the same as for injection from below,
as in Fig.~\ref{fig:model_geometry}(b). In the latter case, electric current bursts
into the ferromagnet like water from a fountain, and flows back through the
normal film: we will call it a ``spin fountain'' device.

We place the origin at the spin injection point, and direct the axes as shown in Fig.~\ref{fig:device}.
%$\hat{x}$ axis along the F/N boundary, the $\hat{y}$ axis transversely to it,
%and the $\hat{z}$ axis normal to the plane of the Fig.~\ref{fig:model_geometry}.
All quantities are assumed to be $z$-independent.
For brevity, we introduce notations $\lambda_n~\equiv~\lambda_s(N)$, $\lambda_f~\equiv~\lambda_s(F)$,
$\sigma_N~\equiv~\sigma(N)$, $\sigma_F~\equiv~\sigma(F)$, and $p~\equiv~p\,(F)$.
For the reasons explained above, we assume $t_N/\lambda_n \ll 1$,
while $t_F/\lambda_f$ can take any value. We switch to a ``mixed potential''
$M~=~\mu~+~p\mu^s / 2$, whereby the bulk equations decouple:
\begin{equation} \label{eq:M_mus_bulk}
\Delta M = 0 \ , \quad
\lambda_f^2 \Delta\mu^s = \mu^s \ .
\end{equation}
The price to pay for this simplification is the change of the boundary
conditions. While $\mu^s$ remains continuous, $M$ experiences a jump
$M_F - M_N = (p/2)\mu^s$ at the F/N interface. Expressions for the currents
now read
\begin{eqnarray}
\label{eq:j_ferro_M-mus}
{\bf j} &=& - \frac{\sigma}{e^2} \vec\nabla M \ ,
 \\
 \label{eq:js_ferro_M-mus}
{\bf j}^s &=& - \frac{\sigma}{e^2} \left( (1-p^2)\vec\nabla\mu^s + 2 p \vec\nabla M \right)
\end{eqnarray}

In a thin normal film, we approximate $M_N(x,y)$, $\mu^s_N(x,y)$ by
their averages over the film thickness $M_N(x)$ and $\mu^s_N(x)$, for which we
derive effective equations
\begin{eqnarray}
\nonumber
R \, \partial^2_x M_N &=& -\frac{1}{t_N} \partial_y M_F (x,0) \ ,
 \\
\label{eq:M_film}
\lambda_n^2 \partial^2_x \mu^s_N &=& \mu^s_N - \frac{\lambda_{mix}^2}{t_N} \times
\\
\nonumber
& \times & \partial_y \left[ \mu^s_F(x,0) + \frac{2 p}{1-p^2} M_F(x,0) \right] - s\delta(x),
\end{eqnarray}
where $R = \sigma_N/\sigma_F$, $\lambda^2_{mix}(p) =  (1-p^2)\lambda_n^2/R$, and $s$ is
a rescaled total injected spin current.

In the ferromagnet, we seek the solutions in the form
\begin{eqnarray}
\nonumber
\mu^s_F(x,y) &=& \int \frac{dk}{2\pi} a_k \cos(k x)
 \frac{\cosh[q(k)(t_F - y)]}{\cosh[q(k) t_F ]} \ ,
 \\
\label{eq:Fourier_M_F_finite_tF}
M_F(x,y) &=& \int \frac{dk}{2\pi} b_k \cos(k x)
 \frac{\cosh[k(t_F - y)]}{\cosh[k t_F ]} \ .
\end{eqnarray}
With $q^2(k) = \lambda_f^{-2} + k^2$, the $\mu^s_F$ and $M_F$ automatically
satisfy the Eqs.~(\ref{eq:M_mus_bulk}) and the boundary conditions
$j_{\perp} = 0$, $j^s_{\perp} = 0$ at the top surface $y = t_F$ of the ferromagnet.

\begin{figure}[t]
\includegraphics[width=0.5\textwidth]{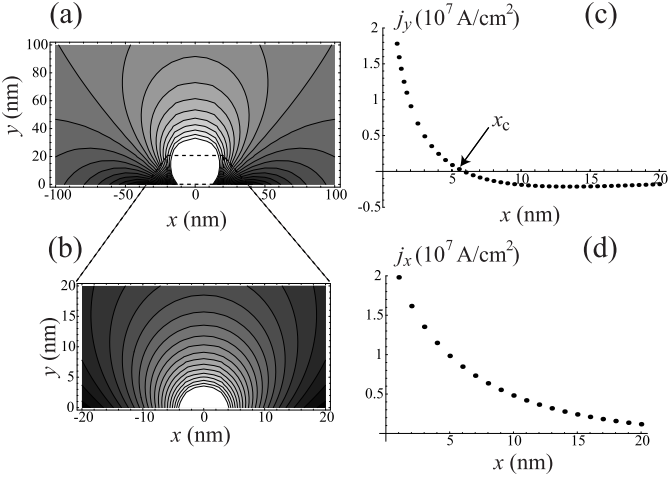}
    \caption{(a) Contour plot of the mixed potential $M_F(x,y)$ for $t_F \to \infty$.
    Electric current $\bf j$ is perpendicular to the contours, Eq.~(\ref{eq:j_ferro_M-mus}).
    (b) Blow-up of (a) near the origin. (c) Normal component $j_y$ of electric current at
    the F/N boundary. The zero of $j_y$ defines the position $x_c$ of the vortex core.
    (d) Component $j_x$ along the F/N boundary, in the F-layer.}
  \label{fig:electric_currents}
\end{figure}

In the normal film $\mu^s_N(x) = \mu^s_F(x,0)$, and $M_N$ is found from the
boundary condition on its jump:
\begin{equation}\label{eq:expansion_MN}
M_N(x) = \int \frac{dk}{2\pi} \left( b_k - \frac{p}{2}a_k \right) \cos(k x) \ .
\end{equation}
Substituting the Fourier expansions into the Eqs.~(\ref{eq:M_film}), we find
the coefficients
\begin{equation}
\label{eq:ak_bk_finite_tF}
a_k = \frac{s}{F(k)} \ ,
\quad
b_k = \frac{p}{2} \ \frac{s H(k)}{F(k)} \ .
\end{equation}
with
\begin{eqnarray}
\nonumber
F(k) &=& f(k) + \frac{p^2}{1-p^2} \frac{\lambda^2_{mix} }{t_N} H(k) k \tanh(t_F k)
\\
\label{eq:fHF}
f(k) &=& 1 + \lambda^2_n k^2 + \frac{\lambda^2_{mix}}{t_N}q(k) \tanh(t_F q(k))
\\
\nonumber
H(k) &=& \frac{Rt_N k }{Rt_N k + \tanh(t_F k)}
\end{eqnarray}
As per Eq. (\ref{eq:ak_bk_finite_tF}), at $p = 0$ the electric current vanishes.

Solutions (\ref{eq:Fourier_M_F_finite_tF}) are computed by numerical
integration. The magnitude of electric current is proportional to the
injected spin current $s$. To compare with experiment, we rescale $s$ so that
spin accumulation $\mu^s_0$ at the injection point has the largest feasible
value, estimated \cite{kimura_prl2006, kimura_prl2007} as $\mu^s_0 \sim 1$
mV. Electric current can be found from (\ref{eq:j_ferro_M-mus}) using
the parameters, typical of a Py/Cu device:\cite{hamrle_prb2005}
$\lambda_n = 350$~nm, $\lambda_f = 4.3$ nm, $t_N = 2$ nm, $p = 0.7$,
$R = 6.6$, and $\sigma(Cu) = 48 \times 10^6$ ($\Omega$m)$^{-1}$.

\begin{figure}[t]
\center
\includegraphics[width=0.5\textwidth]{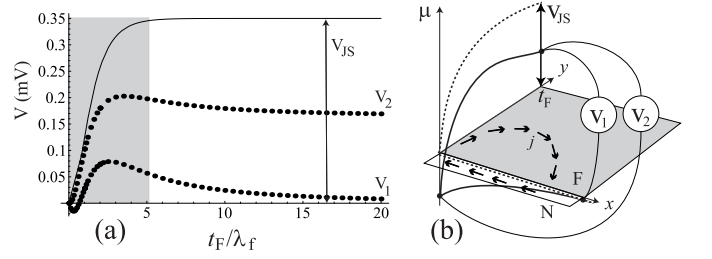}
    \caption{(a) Voltages $V_{1,2}(t_F)$ measured in a spin-fountain device by voltmeters shown in (b). Solid line shows the voltage measured by either of the two voltmeters in a device with uniform spin injection. Gray area marks the region where $\mu_s$ does not fully relax at $y =t_F$. (b) Sketch of electric potential $\mu(x,y)$ in the device. Arrows in the $(x,y)$ plane show the flow of electric current. Solid lines: actual $\mu(x,0)$ and $\mu(0,y)$.  Dotted lines: the same for uniform injection.}
\label{fig:non_local_voltage}
\end{figure}

A typical contour plot of $M(x,y)$ for $t_F \to \infty$ is shown
in the Fig.~\ref{fig:electric_currents}(a,b). Electric current is normal to the 
$M = const$ lines (\ref{eq:j_ferro_M-mus}), and forms a fountain-like pattern
sketched in the Fig.~\ref{fig:model_geometry}(b).
The Figs.~\ref{fig:electric_currents}(c,d) give the current components at the F/N interface.

Induced electric current significantly alters the voltage measured in a Johnson-Silsbee
experiment and makes it dependent on the position of the voltmeter probe.
Let us assume that the F-probe is attached at the top of the F-layer, right
above the injection point. For an extended F-electrode, the easiest way to
attach an N-probe is at $x_N \to \infty$ (Fig.~\ref{fig:non_local_voltage},
b). Then the measured voltage $V_1(t_F) = \mu(0,t_F) - \mu(x_N,0) \to
\mu(0,t_F)$. The plot of $V_1(t_F)$ is given in
Fig.~\ref{fig:non_local_voltage}(a). If the N-probe is attached close to the
spin-injection point, the voltage changes to $V_2$. Both $V_1$ and $V_2$
significantly differ from the voltage that would develop in the absence of
electric current (solid line in Fig.~\ref{fig:non_local_voltage}-a).
Dependence on the F-electrode thickness is also quite visible even for $t_F
\gg \lambda_f$. The $j = 0$ situation emerges either in a narrow F-electrode,
or, more generally, in devices where spin is injected uniformly across the
F/N interface: According to the Eq.~(\ref{eq:curl_j}), when $\nabla \mu^s$ is
normal to the boundary, the reason for current generation vanishes together
with ${\rm curl} \, {\bf j}$. Uniform spin injection generates voltage that
approaches the Johnson and Silsbee result $V_{\rm JS} = (p/2) \mu^s_0 = 0.35$ mV
for $y \gg \lambda_f$.

Spin current tends to decay exponentially with the distance
from the injection point. For example, for a non-magnetic  top
layer, in the present case of $R t_N \gg \lambda_f$, we find $a_k \approx
1/(f(0) + \lambda_n^2 k^2)$. Hence, along the interface the spin
potential falls off as $\mu^s(x,0) \approx \mu^s_0 \exp{(-x/\lambda_{\|})}$.
The decay length $\lambda_{\|} = \lambda_n/\sqrt{f(0)}$ is bound as per
$\lambda_f < \lambda_{\|} < \lambda_n$. Physically, this means that spins
would propagate through a detached normal metal film up to a length of
about $\lambda_n$, but the spin current leakage into the overlayer
shortens their reach.

\begin{figure}[t]
\includegraphics[width=0.45\textwidth]{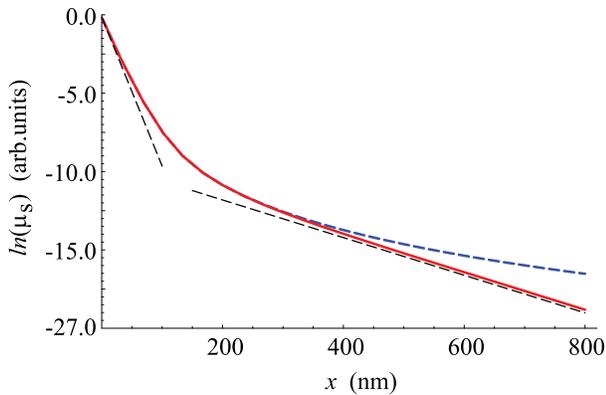}
\caption{(Color online) Log plot of spin potential. Red, solid line: $\mu^s(x,0)$ for $t_F = 250$ nm. Blue, dashed line: $\mu^s(x,0)$ for $t_F \to \infty$. Dashed linear fits: short-range exponential decay with $\lambda_{\|} = 10.5$ nm and long-range exponential decay with $\lambda_c \approx 83$ nm.}
\label{fig:mus_of x}
\end{figure}

For a magnetic top layer ($p \neq 0$), a log plot of $\mu^s(x,0)$ is shown in
Fig.~\ref{fig:mus_of x}: the $\mu^s(x,0)$ decays exponentially. However, the
decay length crosses over from $\lambda_{\|}$ at $x \ll t_F$ to a longer
length $\lambda_c$ at $x \gg t_F$.
In the thick-film limit, $R t_N/t_F \ll 1$, we find % were able to obtain an approximation
$$
\lambda_c = \frac{t_F}{\pi} \left(1 + \frac{R t_N}{t_F} + \ldots \right) \ .
$$
For the parameters above, this yields $\lambda_c \approx 84$
nm against the numerically found $\lambda_c \approx 83$ nm.

The $\lambda_c$ grows with $t_F$ and, at $t_F \to \infty$, the $\mu^s(x,0)$
decays non-exponentially for $x \gg \lambda_{\|}$. In this limit, $\tanh(t_F
k) \to {\rm sgn}(k)$ is non-analytic at $k=0$, and the expressions
for $F(k)$ and $H(k)$ read
\begin{equation}\label{eq:fHF_infinite_tF}
F(k) = f(k) + \frac{p^2}{1-p^2} \frac{\lambda^2_{mix} }{t_N} H(k) |k|
 , \
H(k) = \frac{Rt_N |k| }{Rt_N |k| + 1} \ .
\end{equation}
The singularity shows itself as a $|k|^3$ term in the expansion of $a_k$.
Using the stationary phase method, we find an asymptotic expression
\begin{equation}
\label{eq:asymptotic_FN_infinite_tF}
\mu^s(x,0) \sim \frac{C}{x^4} + \ldots \qquad (|x| \gg \lambda_{\|})
\end{equation}
with $C = s(p^2/(1-p^2)) (6 R^2/\pi f^2(0)) \lambda_{mix}^2 t_N$.
Thus, for infinite $t_F$, the spin accumulation ultimately decays
as a power-law, i.e., very slowly (the blue line in the Fig.~\ref{fig:mus_of x}).
%Thus, in the $t_F \to \infty$ limit, spin accumulation decays
%as a power-law, i.e., very slowly at $x \gg \lambda_\|$
%(the blue line in the Fig.~\ref{fig:mus_of x}).

Our findings mean that a ferromagnetic overlayer makes the injected
spin current propagate further along the normal film. This conclusion sounds
pronouncedly counter-intuitive: After all, ferromagnetic layer is known to be a
spin sink, so one would expect that it could only lower the spin propagation
length. The seeming paradox is resolved as follows. As we know, the electric
current loops cross the F/N boundary. Upon such crossing, a non-equilibrium
spin density is inevitably produced \cite{vanson_prl1987}, so $\mu^s$
cannot decay independently of $\bf j$. Ultimately, the conservation of $\bf j$,
expressed by the first equation (\ref{eq:divj_divjs}), causes
a long-range propagation of both charge and spin.
The current-assisted propagation of spin also explains the role of the
F-layer thickness. The long-range pattern of $\bf j$ is limited by the outer
boundaries of the device. Finite $t_F$ is equivalent to
``covering the fountain by a lid'', deflecting $\bf j$ down to the normal film
within a distance of the order of $t_F$. Beyond this distance, the
power-law decay of spin accumulation reverts to the exponential form.

In conclusion, we have shown that the gradient of spin accumulation along an F/N interface produces closed electric current loops. The first consequence of this is a significant reduction of the Johnson-Silsbee voltage,
which means that the interpretation of some non-local resistance experiments
with wide F-electrodes may need to be revisited. For example, in the Refs.~\onlinecite{jedema_nature2001, jedema_prb2003, kimura_jpcm2007, erekhinsky_apl2010}, the current polarization of permalloy was deduced to be
$p \lesssim 0.3$, which is significantly smaller than $p \approx 0.7$ inferred from
GMR measurements.\cite{steenwyk_jmmm1997, holody_prb1998, dubois_prb1999, hamrle_prb2005} Such a seeming reduction of $p$ may arise due to loop currents; this can be verified  by voltage measurements on a series of devices with varying thickness or width of the F-electrodes.

Alternatively, loop currents will manifest themselves by non-zero voltage
drop along the normal film. In the absence of electric current,
such a voltage must vanish --- but not if ${\bf j} \neq 0$. In particular,
if one probe is connected at $x_N \to \infty$ and another at $x_N = 0$,
the voltmeter will read off the voltage difference $V_2 - V_1$,
shown in Fig.~\ref{fig:non_local_voltage}.

Another consequence of loop currents is the long-range propagation of spin
accumulation along the F/N interface. This effect can be measured by an
additional F-electrode positioned downstream of the wide F-electrode. The
non-local voltage on the former will reflect the enhanced propagation of spins
brought about by the latter. Notice that the signal to be expected in such an
experiment is small: As shown in Fig.~\ref{fig:mus_of x}, spin accumulation
drops by orders of magnitude before the long-range propagation regime
becomes sufficiently pronounced.

More generally, electric current vortices at the interface between two materials
shall be expected whenever electric current is coupled to another driven diffusive
current by linear relationships with material-dependent Onsager cross-coefficients.
For example, coupling with heat flow may induce electric current loops in
spin-caloritronic devices with a temperature gradient along the interface between
detector ferromagnet and a normal wire.\cite{bakker_prl2010, slachter_natphys2010, erekhinsky_apl2012}

Ya.\ B. was supported by the NSF grant DMR-0847159. He is grateful to Laboratoire de Physique Th\'eorique, Toulouse, for the hospitality, and to CNRS for funding the visits.


\begin{thebibliography}{2}
\bibitem{maekawa_book2012}
{\em Spin current}, Eds. S. Maekawa, S. O. Valenzuela, E. Saitoh, and T. Kimura, Oxford University Press (2012).

\bibitem{johnson-silsbee_prl1985}
M. Johnson and R. H. Silsbee,
{\em Interfacial charge-spin coupling: Injection
and detection of spin magnetization in metals},
Phys. Rev. Lett. {\bf 55} 1790 (1985).

\bibitem{vanson_prl1987}
P. C. van Son, H. van Kempen, and P. Wyder,
{\em Boundary resistance of the ferromagnetic-normal metal interface},
Phys. Rev. Lett. {\bf 58}, 2271 (1987).

\bibitem{valet-fert_prb1993}
T. Valet and A. Fert,
{\em Theory of the perpendicular magnetoresistance in magnetic multilayers},
Phys. Rev. B {\bf 48}, 7099 (1993).

\bibitem{takahashi_prb2003}
S. Takahashi and S. Maekawa,
{\em Spin injection and detection in magnetic nanostructures},
Phys. Rev. B {\bf 67}, 052409 (2003).

\bibitem{rashba_prbrc2000}
E. I. Rashba,
{\em Theory of electrical spin injection: Tunnel contacts as a solution
of the conductivity mismatch problem},
Phys. Rev. B {\bf 62}, R16267 (2000).

\bibitem{rashba_epjb2002}
E. I. Rashba,
{\em Diffusion theory of spin injection through resistive contacts},
Eur. Phys. J. B {\bf 29} 513 (2002).

\bibitem{hamrle_prb2005}
J. Hamrle, T. Kimura, Y. Otani, K. Tsukagoshi, and Y. Aoyagi,
{\em Current distribution inside Py/Cu lateral spin-valve devices},
Phys. Rev. B {\bf 71}, 094402 (2005).

\bibitem{note} In the Refs.~\onlinecite{hamrle_prb2005, johnson_prb2007, nakane_ieeeletters2012}, they do.

\bibitem{johnson_prb2007}
M. Johnson and R. H. Silsbee,
{\em Calculation of nonlocal baseline resistance in a quasi-one-dimensional wire},
Phys. Rev. B {\bf 76}, 153107 (2007).

\bibitem{nakane_ieeeletters2012}
R. Nakane, S. Sato, S. Kokutani, and M. Tanaka,
{\em Appearance of Anisotropic Magnetoresistance and Electric Potential Distribution in
Si-Based Multiterminal Devices With Fe Electrodes},
IEEE Magn. Lett. {\bf 3},  3000404 (2012).

\bibitem{footnote_number_currents}
Currents ${\bf j}_{\sigma}$ are defined here as particle number currents. Electric currents are obtained by multiplying by electron charge.

\bibitem{kimura_prl2006}
T. Kimura, Y. Otani, and J. Hamrle,
{\em Switching magnetization of nanoscale ferromagnetic particle using nonlocal spin injection},
Phys. Rev. Lett. {\bf 96}, 037201 (2006).

\bibitem{kimura_prl2007}
T. Kimura, Y. Otani,
{\em Large spin accumulation in permalloy-silver lateral spin valve},
Phys. Rev. Lett. {\bf 99}, 196604 (2007).

% low P measurements in NLSV
% {jedema_nature2001, jedema_prb2003, kimura_jpcm2007, erekhinsky_apl2010}
\bibitem{jedema_nature2001}
F. J. Jedema, A. T. Filip, and B. J. van Wees,
{\em Electrical spin injection and accumulation at room temperature in an all-metal mesoscopic spin valve}
Nature {\bf 410}, 345 (2001).

\bibitem{jedema_prb2003}
F. J. Jedema, M. S. Nijboer, A. T. Filip, and B. J. van Wees,
{\em Spin injection and spin accumulation in all-metal mesoscopic spin valves}
Phys. Rev. B {\bf 67}, 085319 (2003).

\bibitem{kimura_jpcm2007}
T. Kimura and Y. Otani,
{\em Spin transport in lateral ferromagnetic/nonmagnetic hybrid structures},
J. Phys.: Condens. Matter {\bf 19}, 165216 (2007).

\bibitem{erekhinsky_apl2010}
%Mikhail Erekhinsky, Amos Sharoni, Felix Casanova, and Ivan K. Schuller,
M. Erekhinsky, A. Sharoni, F. Casanova, and I. K. Schuller,
{\em Surface enhanced spin-flip scattering in lateral spin valves},
Appl. Phys. Lett. {\bf 96}, 022513 (2010).

% high P measurements in GMR
% {steenwyk_jmmm1997, holody_prb1998, dubois_prb1999, }
\bibitem{steenwyk_jmmm1997}
S. D. Steenwyk, S. Y. Hsu, R. Loloee, J. Bass, and W. P. Pratt, Jr.
{\em Perpendicular-current exchange-biased spin-valve evidence for a short spin-diffusion length in permalloy},
J. Magn. Magn. Mater. {\bf 170}, L1 (1997).

\bibitem{holody_prb1998}
P. Holody, W. C. Chiang, R. Loloee, J. Bass, W. P. Pratt, Jr., and P. A. Schroeder,
{\em Giant magnetoresistance of copper/permalloy multilayers},
Phys. Rev. B {\bf 58}, 12230 (1998).

\bibitem{dubois_prb1999}
S. Dubois, L. Piraux, J. M. George, K. Ounadjela, J. L. Duvail, and A. Fert,
{\em Evidence for a short spin diffusion length in permalloy from the giant magnetoresistance of multilayered nanowires},
Phys. Rev. B {\bf 60} 477 (1999).

% Caloritronics
%{bakker_prl2010, slachter_natphys2010, erekhinsky_apl2012}

\bibitem{bakker_prl2010}
F. L. Bakker, A. Slachter, J.-P. Adam, and B. J. van Wees,
{\em Interplay of Peltier and Seebeck Effects in Nanoscale Nonlocal Spin Valves},
Phys. Rev. Lett. {\bf 105}, 136601 (2010).

\bibitem{slachter_natphys2010}
A. Slachter, F. L. Bakker, J-P. Adam, and B. J. vanWees,
{\em Thermally driven spin injection from a ferromagnet into a non-magnetic metal},
Nat. Phys. {\bf 6}, 879 (2010).


\bibitem{erekhinsky_apl2012}
M. Erekhinsky, F. Casanova, I. K. Schuller, and A. Sharoni
{\em Spin-dependent Seebeck effect in non-local spin valve devices},
Appl. Phys. Lett. {\bf 100}, 212401 (2012).

\end{thebibliography}
\end{document}